\documentclass{PoS}

\title{Nucleon spectroscopy using multi-particle operators}

\ShortTitle{Nucleon spectroscopy using multi-particle operators}

\author{\speaker{Waseem Kamleh}%
\thanks{This research was undertaken with the assistance of resources
  at the NCI National Facility in Canberra, Australia, and the iVEC
  facilities at the University of Western Australia (iVEC@UWA),
  provided through the National Computational Merit Allocation Scheme
  and the University of Adelaide Partner Share. We also acknowledge
  the support of eResearch SA. This research is supported by the
  Australian Research Council.} \\
        Special Research Centre for the Subatomic Structure of Matter,\\
        University of Adelaide, Australia\\ 
        E-mail: \email{waseem.kamleh@adelaide.edu.au}}

\author{Adrian L. Kiratidis \\
       Special Research Centre for the Subatomic Structure of  Matter, \\
       University of Adelaide, Australia}

\author{Derek B. Leinweber \\
       Special Research Centre for the Subatomic Structure of  Matter, \\
       University of Adelaide, Australia}

\abstract{The role of 5-quark operators in extracting the nucleon
  excited state spectrum via correlation matrix techniques is
  explored. In particular, the coupling of meson-baryon operators to
  nucleon resonance states and scattering states is studied. Results
  are presented for 2+1 flavour dynamical ensembles in both the
  positive and negative parity channels. Fitting a single-state ansatz
  to the eigenstate-projected correlators provides robust energies for
  the low-lying spectrum that are essentially invariant across a
  variety of different operator bases. In particular, the resonant
  state energies obtained both with and without the use of
  meson-baryon operators agree, demonstrating that resonance
  energies can be reliably extracted solely using 3-quark operators.}

\FullConference{The 32nd International Symposium on Lattice Field Theory\\
		 23-28 June, 2014\\
		 Columbia University New York, NY}

\usepackage{graphicx}
\usepackage{amsmath}

\begin{document}

\section{Introduction}

The use of correlation matrix techniques~\cite{Michael:1985ne,
  Luscher:1990ck} to study the nucleon excited state spectrum has seen
a significant amount of recent activity, with lattice studies being
performed in both the positive parity~\cite{Roberts:2013ipa,
  Edwards:2011jj, Mahbub:2009aa, Mahbub:2010rm, Liu:2014jua} and
negative parity~\cite{Edwards:2011jj, Lang:2012db, Mahbub:2012ri, Mahbub:2013bba}
sectors.
The underlying principle is to begin with a sufficiently large basis
of $N$ operators (so as to span the space of the states of interest
within the spectrum) and construct an $N \times N$ matrix of cross
correlation functions,
\begin{equation} {G_{ij}(t,\vec p)}=\sum_{\vec x}e^{-i{\vec p}.{\vec x}}\langle{\Omega}\vert T \{ \chi_i (x)\bar\chi_j(0)\} \vert{\Omega}\rangle. \end{equation}
We then solve a generalised eigenproblem 
to find the linear combination of interpolating fields,
\begin{align}  
{\bar\phi}^{\alpha} =\sum_{i=1}^{N} u_{i}^{\alpha}\, {\bar\chi}_{i}, & &  {\phi}^{\alpha} =\sum_{i=1}^{N} v_{i}^{\alpha}\, {\chi}_{i}
\end{align}
such that $\phi^{\alpha}$ and $\bar{\phi}^{\alpha}$ couple to a single energy eigenstate (labelled by $\alpha$) and the correlation matrix is diagonalised,
\begin{equation} v_{i}^{\alpha}G_{ij}(t)u_{j}^{\beta} = \delta^{\alpha\beta}z^{\alpha}{\bar{z}}^{\beta}e^{-m_{\alpha}t}. \end{equation}
The left and right vectors are then used to define the eigenstate-projected correlators (after applying the parity projection operators at $\vec{p}=0$),
\begin{align}
 v_{i}^{\alpha}G^{\pm}_{ij}(t)u_{j}^{\alpha} & \equiv G^{\alpha}_{\pm}(t).
\end{align}
At this point we note that if the operator basis does
not appropriately span the low-lying spectrum $G^{\alpha}_{\pm}(t)$
may contain a mixture of two or more energy eigenstates. There are a
number of scenarios in which this might occur:
\begin{itemize}
\item At early Euclidean times the number of states strongly
  contributing to the correlation matrix may be (much) larger than the
  number of operators in the basis.

\item There may be energy eigenstates present that do not couple or
  only couple weakly to the operators used. In particular, it is well
  known that local 3-quark interpolating fields couple poorly to
  multi-hadron scattering states.

\item The nature of the operators selected may be such that it is not
  possible to construct a linear combination with the appropriate
  structure to isolate a particular state.
\end{itemize}
It is important to have a strategy to ensure that one can
accurately obtain eigenstate energies from the correlation matrix.
Now, even if a mixture of states is present, asymptotically Euclidean
time evolution will tend to isolate a single state,
\begin{align}
G^{\alpha}_{\pm}(t) & \stackrel{t \rightarrow \infty}{=} z_\alpha \bar{z}_\alpha e^{-m_\alpha t} .
\end{align}
Hence, the method we use is to construct effective masses of different
states from the eigenstate-projected correlators and then analyse them
in the usual way. As we will demonstrate, a careful $\chi^2$ analysis
to fit the single-state ansatz ensures a robust extraction of the
eigenstate energies.

As described above, the choice of an appropriate operator basis is
critical to obtaining the complete spectrum of low-lying excited
states. Recall that we can expand any radial function using a basis of
Gaussians of different widths $f(|\vec{r}|) = \sum_i c_i
e^{-\varepsilon_i r^2}.$ This leads to the use of Gaussian-smeared
fermion sources with a variety of widths~\cite{Burch:2004he}, providing an
operator basis that is highly suited to accessing radial
excitations. Indeed, the CSSM lattice collaboration has used this well
established technique~\cite{Mahbub:2010rm,Mahbub:2012ri} in the
calculation of the positive and negative parity nucleon excited state
spectra.  It is the combination of Gaussian sources of different
widths that allows for the formation of the nodal structures needed to
isolate the different radial excitations~\cite{Roberts:2013oea}.

However, it has been shown that a basis consisting solely of
three-quark operators has difficulty in detecting multi-particle
scattering energy levels~\cite{Mahbub:2013bba}.  The coupling of
three-quark baryon operators to multi-particle states is strongly
suppressed, leading us to consider the inclusion of interpolating
fields that we expect to have substantial overlap with multi-particle
meson-baryon type states~\cite{Lang:2012db}.  One possible solution is
to explicitly include hadron-hadron type interpolators, as has been
done with mesons~\cite{Morningstar:2013bda} by combining single-hadron
operators with the relevant momentum.  This creates an operator that
necessarily has a high overlap with the scattering state of interest,
enabling its extraction.

In this work we take a different approach, and instead aim to
construct meson-baryon type interpolators without explicitly
projecting momenta to investigate the role that the resulting operator
plays in the calculation of the nucleon spectrum.
\begin{figure}[t]
  \centering
  {\includegraphics[width=0.6\textwidth]{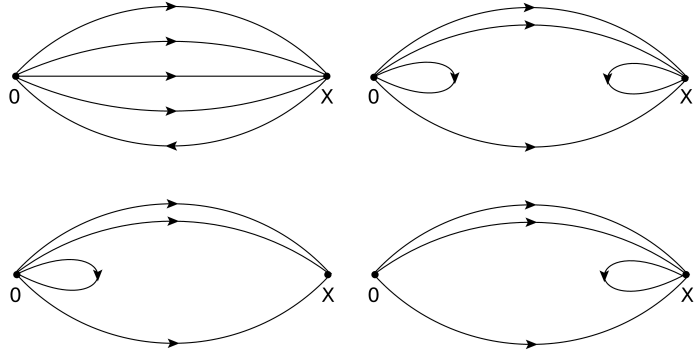}}
 \caption{The Feynmann diagrams present following the introduction of five-quark interpolating fields.}
  \label{fig:FeynmannDiag}
\end{figure}
Using these operators we perform spectroscopic calculations using a
variety of bases containing both three- and five-quark
operators. Examining the resulting spectra provides us with an
excellent opportunity to test the robustness of the variational
techniques employed.

\section{Simulation Details}

For this work we use the PACS-CS $2 + 1$ flavour dynamical-fermion configurations~\cite{Aoki:2008sm} made available through the ILDG~\cite{Beckett:2009cb}.  These configurations use the non-perturbatively $\mathcal{O}(a)$-improved Wilson fermion action and the Iwasaki gauge action.  The lattice size is $32^3 \times 64$ with a lattice spacing of $0.0907\textrm{ fm}$ providing a physical volume of $\approx (2.90\textrm{ fm})^3$.  $\beta = 1.90$, the light quark mass is set by the hopping parameter $\kappa_{ud} = 0.13770$ which gives a pion mass of $m_{\pi} = 293 \textrm{ MeV}$, while the strange quark mass is set by $\kappa_{s} = 0.13640$.  
Gauge-invariant Gaussian smearing~\cite{Gusken:1989qx} is employed at the source and sink which alters the overlap of the operator with the various states in the spectrum.
The source is inserted at $t_s = 16$, while the variational analysis is done at $t_{0}=17$ and $dt=3$ which provides a good balance between systematic and statistical uncertainties.  Uncertainties are obtained via single elimination jackknife while a full covariance matrix analysis provides  $\chi^{2}/dof$ which is utilized to select fits.

We make use of the following conventional three-quark nucleon operators,
\begin{align*}
\chi_{1} &= \epsilon^{abc}[u^{aT}\, (C\gamma_{5})\, d^{b}]\, u^{c},\\
\chi_{2} &= \epsilon^{abc}[u^{aT}\, (C)\, d^{b}]\, \gamma_{5}\, u^{c}.
\end{align*}
Utilising the Clebsch-Gordon coefficients to project to $I = 1/2,
I_{3} = +1/2$, we can write down the general form of our meson-baryon
interpolating fields~\cite{Kiratidis:2012mr},
\begin{align}\label{Proton5QrkOpInterpolator}
\chi_{N \pi}(x) 
= \frac{1}{2\sqrt{3}}\,\epsilon^{abc}
\bigg\{2&\big(u^{Ta}(x)\,\Gamma_{1}\,d^{b}(x)\big)\,\Gamma_{2}d^{c}(x)\,\Big[\bar{d}^{e}(x)\,\gamma_{5}\,u^{e}(x)\Big]\nonumber\\
- &\big(u^{Ta}(x)\,\Gamma_{1}\,d^{b}(x)\big)\,\Gamma_{2}u^{c}(x)\,\Big[\bar{d}^{e}(x)\,\gamma_{5}\,d^{e}(x)\,\Big]\nonumber\\
+ &\big(u^{Ta}(x)\,\Gamma_{1}\,d^{b}(x)\big)\,\Gamma_{2}u^{c}(x)\,\Big[\,\bar{u}(x)^{e}\,\gamma_{5}\,u^{e}(x)\Big]\bigg\},
\end{align}
providing us with two five-quark operators, denoted $\chi_{5}$ and
$\chi^{\prime}_{5}$ which correspond to $(\Gamma_{1}, \Gamma_{2}) =
(C\gamma_{5}, \textrm{I})$ and $(\Gamma_{1}, \Gamma_{2}) =
(C,\gamma_{5})$ respectively. Note, these transform negatively under
parity, so we multiply our five-quark operators by $\gamma_5$ before
constructing the correlation matrix.

We form seven operator bases by selecting different combinations of the four nucleon interpolating fields as outlined in the following table: 
\begin{table}[ht]
\centering
\begin{tabular}{cc}
 \hline
\noalign{\vspace{3pt}}
    Basis Number & Operators Used  \\
\noalign{\vspace{3pt}}
    \hline 
\noalign{\vspace{3pt}}
    1 & $\chi_{1}$, $\chi_{2}$\\
    2 & $\chi_{1}$, $\chi_{2}$, $\chi_{5}$\\
    3 & $\chi_{1}$, $\chi_{2}$, $\chi_{5}^{\prime}$\\
    4 & $\chi_{1}$, $\chi_{2}$, $\chi_{5}$, $\chi_{5}^{\prime}$\\
    5 & $\chi_{1}$, $\chi_{5}$, $\chi_{5}^{\prime}$\\
    6 & $\chi_{2}$, $\chi_{5}$, $\chi_{5}^{\prime}$\\
    7 & $\chi_{5}$, $\chi_{5}^{\prime}$\\
\hline 
\end{tabular}
\label{table:BasisTable}
\end{table}

The presence of creation quark fields in our annihilation
interpolating field and vice-versa leads to the requirement of
calculating the loop propagators, present in the right-hand diagram of
Figure \ref{fig:FeynmannDiag}.  There are a number of ways of dealing
with such diagrams including distillation~\cite{Peardon:2009gh}, and
the Laplacian Heaviside (LapH) smearing method~\cite{Morningstar:2011ka}.  Here we will stochastically estimate
inverse matrix elements fully diluting in spin, colour and time.

Throughout this work we employ two different levels of fermion source
and sink smearing to increase the basis size and provide access to
radial excitations. We use $n_s = 35$ and $n_s = 200$ sweeps of smearing
providing bases of sizes 4, 6 and 8.

\section{Results}
\label{sect:Results}

\begin{figure}[!th]
  \centering 
  {\includegraphics[width=0.7\textwidth]{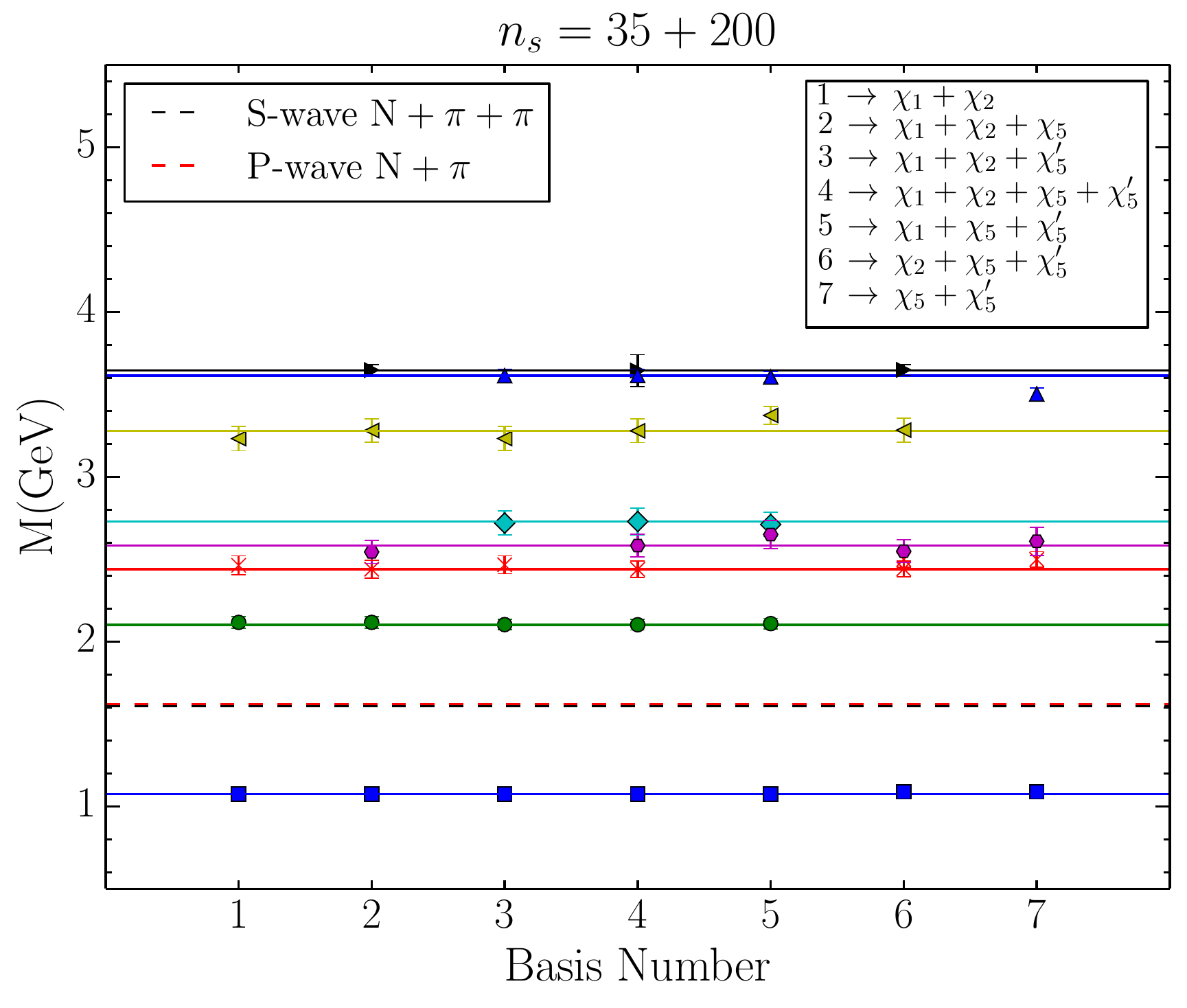}}
 \caption{Positive parity nucleon spectrum with various operator bases
   with 35 and 200 sweeps of smearing.  Horizontal solid lines are
   present to guide the eye and are drawn from the central value of
   the states in basis 4, since this basis is the largest.}
 \label{fig:Masses+}
\end{figure}

We now present the nucleon spectrum in the positive parity sector in
Figure \ref{fig:Masses+}.  The solid horizontal lines therein have
been added to guide the eye, having their values set by the states in
basis number 4, since this basis contains all the operators studied
and hence possess the largest span.  

Of particular interest is the robustness of the variational techniques
employed.  While changing bases may effect whether or not a particular
state is seen, the energy of the extracted states is consistent across
the different bases, even though they contain qualitatively different
operators.

The ground-state nucleon is observed in every basis regardless of the
absence or presence of a particular operator.  In contrast, we see
that $\chi_{1}$ is critical to the extraction of the first excited
state, a radial excitation of the ground
state~\cite{Roberts:2013oea}. We see this state in bases 1 through 5,
whereas bases 6 and 7 which lack $\chi_{1}$ do not observe this state.

Despite the use of 5-quark operators, states consistent with the
P-wave $N\pi$ or S-wave $N\pi\pi$ scattering thresholds are absent.
This is understood by noting that none of our operators have a source
of the back-to-back relative momentum necessary to observe the
scattering states in this channel.

\begin{figure}[!th]
  \centering
  {\includegraphics[width=0.7\textwidth]{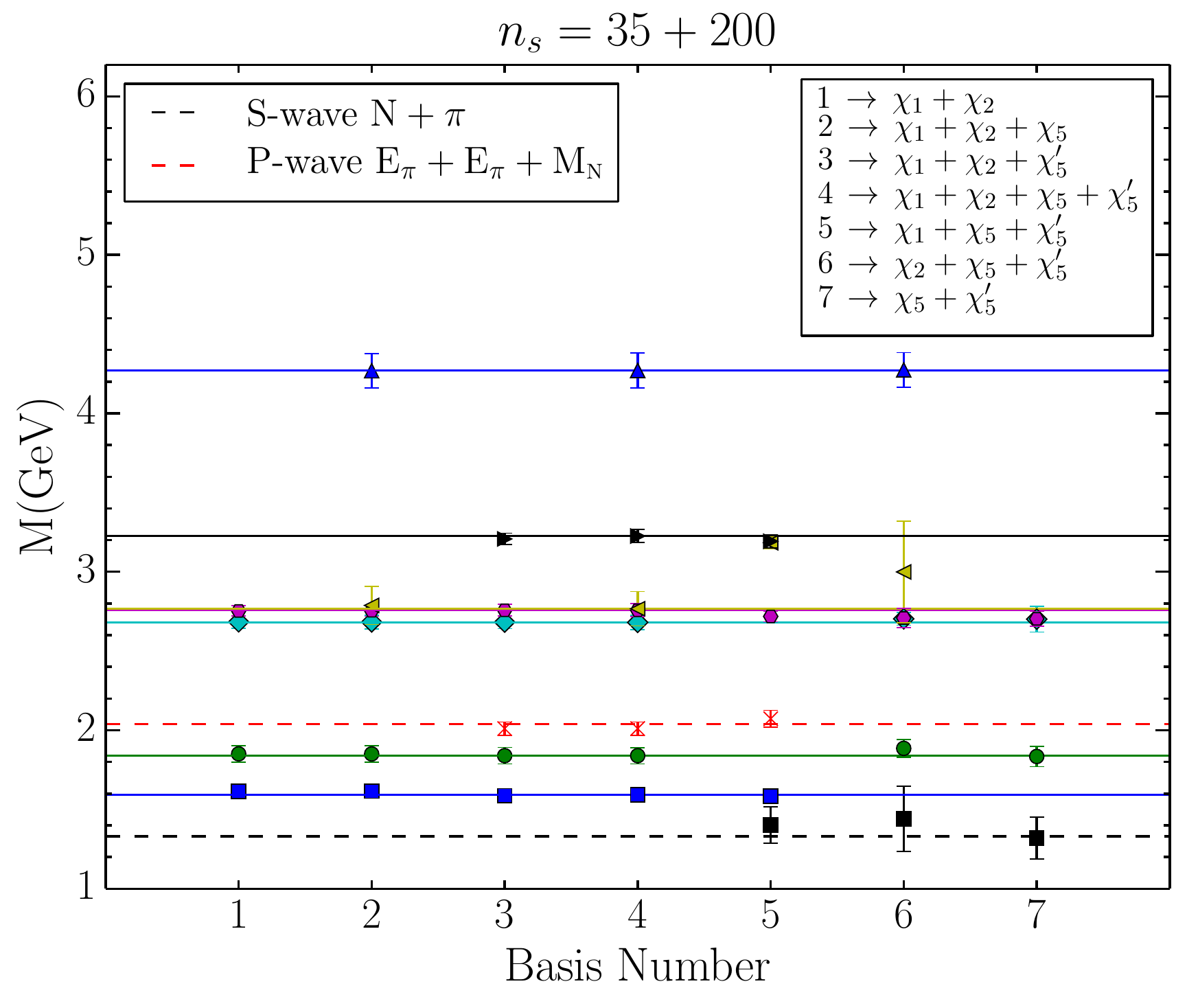}}
 \caption{The negative parity nucleon spectrum with various operator
   bases using 35 and 200 sweeps of smearing.  Solid horizontal lines
   are present to guide the eye and are drawn from the central value
   of the states in basis 4, since this basis is the largest.  The
   variational parameters used herein are $(t_{0},dt) = (17,3)$.}
 \label{fig:Masses-}
\end{figure}

The negative parity nucleon spectrum is presented in Figure
\ref{fig:Masses-}.  Again solid horizontal lines have been added to
guide the eye, having their values set by the states in basis number
4. Once again, while changing bases effects whether or not
we observe a given state, the extracted states display an impressive
level of consistency.
In accord with previous studies~\cite{Mahbub:2012ri, Mahbub:2013ala},
we find that the $\chi_{1}$ interpolating field is crucial for
extracting the lowest negative parity resonance, associated with the
$S_{11}(1535),$ as we do not observe this state when $\chi_{1}$ is absent
as in bases 6 and 7.

In contrast to the positive parity results, we do observe states that
are consistent with the scattering thresholds in the negative parity
channel, specifically the S-wave $N\pi$ (bases 5,6,7) and P-wave $N\pi\pi$ (bases 3,4,5) thresholds. 
It is somewhat surprising that basis 4 fails to see the S-wave $N\pi$
scattering threshold, despite being the largest.  It seems that the
coupling to this scattering state is low relative to the large spectral
strength of the resonant states.
We observe that a scattering state is seen if (and only if) the $\chi_5'$ operator
is included, indicating that the presence of the vector di-quark in
the interpolator may play a significant role in scattering state
excitation.

It is important to note that even after the introduction of operators
that permit access to low-lying scattering states in the spectrum
(such as in bases 5, 6 and 7) the energies of the higher states in the
spectrum are consistent, demonstrating the robustness of the
variational techniques employed.

\section{Conclusion}

We studied the excited state nucleon spectrum on a variety of bases
including both standard 3-quark operators and local multi-particle
operators. We find that fitting a single-state ansatz to the effective
masses of the eigenstate projected correlators results in a robust
extraction of the spectrum in both the positive and negative parity
channels. While the selection of states that were observed varied
between bases, when a given state was seen the extracted energy agreed
across qualitatively different bases.

An important feature of our negative parity results is that the
energies of the resonance states extracted are consistent across all
bases in which the state is observed, regardless of the presence (or
not) of a lower-lying scattering threshold state in the correlation
matrix analysis. This study demonstrates that (by using the techniques
described) one does not need to have access to the scattering states
to reliably extract resonant state energies.


\providecommand{\href}[2]{#2}\begingroup\raggedright\endgroup

\end{document}